# Surface properties of large TNOs: Expanding the study to longer wavelengths with the James Webb Space Telescope

Noemí Pinilla-Alonso[1], John A. Stansberry[3], Bryan J. Holler[3]

[1]Florida Space Institute, UCF, Orlando, FL, USA

[2]Arecibo Observatory, Arecibo, PR

[3]Space Telescope Science Institute, Baltimore, MD, USA

**Abstract**

The largest trans-Neptunian objects (TNOs) represent an extremely diverse collection of primitive bodies in the outer solar system. The community typically refers to these objects as "dwarf planets," though the IAU acknowledges only four TNOs officially as such: Pluto, Eris, Makemake, and Haumea. We present a list of 36 potential candidates for reclassification as dwarf planets, namely candidate dwarf planets (CDPs), which cover a wide range of sizes, geometric albedos, surface colors and probably, composition. Understanding the properties across this population, and how those properties change with size, will yield useful constraints on the environment in which these TNOs formed, as well as their dynamical evolution, and bulk interior composition. TNO surface characteristics are ideal for study with the *James Webb Space Telescope* (*JWST*), which provides imaging and spectroscopic capabilities from ~0.6-28 **µm**. The four available science instruments, MIRI, NIRCam, NIRISS, and NIRSpec, and their capabilities for the study of TNOs, are presented. *JWST* will expand on the wavelength range observable from the ground in the near-infrared (0.6-5 µm) for compositional studies and will open a new window on TNOs in the mid-infrared (5-28 µm) for thermal characterization.

## 1   Introduction

The year 2005 saw the unprecedented discovery of three of the largest known trans-Neptunian objects (TNOs): (136108) Haumea, (136472) Makemake, and (136199) Eris (Brown et al., 2005a; Santos-Sanz et al. 2005). One year later, the International Astronomical Union (IAU) revisited the definition of a planet and introduced a new category of objects in the solar system, the "dwarf planets" (DPs). According to this definition, dwarf planets are small bodies that orbit the Sun and are large enough to be in hydrostatic equilibrium but not large enough to have cleared their orbit of other minor bodies (see Resolution B5 and B6 of the XXVI General Assembly of the IAU at https://www.iau.org/static/resolutions/Resolution_GA26-5-6.pdf). Hence, the fact that dwarf planets and planets fall under different categories is a consequence of different evolutionary processes and a reflection of different physical and dynamical characteristics. In subsequent years, dedicated surveys (e.g. Brown et al., 2015) were used to search for additional TNOs with sizes similar to those of the known DPs. However, these searches, which proved to be 100% effective in detecting bright objects (V≤19) beyond 25 AU, did not identify any new dwarf planet. Consequently, the probability of finding one or two of these objects in the galactic plane, where the stellar background makes detections more difficult, is below 35%. The list of dwarf planets recognized by the IAU has not changed since 2006, even though a handful of large TNOs are unofficially considered to be in this category.

In this work, we review the physical properties of these large TNOs to search for distinctive characteristics that could help to identify new DPs among the larger population. For that, we use the most recent results, which were not available in 2006 when the definition was published. We also discuss how new instrumentation, such as that on board the soon-to-be-launched *James Webb Space Telescope* (*JWST*), can expand our knowledge of the surface properties of dwarf planets and candidates, as well as other TNOs.

## 2    Dwarf planets and candidate dwarf planets

According to the IAU definition, a large TNO must be in hydrostatic equilibrium to be considered a dwarf planet. It is difficult to determine if an object is in hydrostatic equilibrium using modern instrumentation; however, this property is related, among other factors, to the shape of the body. A combination of sufficient internal pressure caused by the body's gravitation, and sufficient plasticity to allow gravitational relaxation would cause bodies larger than a certain size to overcome their internal strength and adopt a shape characteristic of a body in hydrostatic equilibrium, typically a triaxial ellipsoid (e.g., Chandrasekhar, 1987). On the other end of the size spectrum, smaller TNOs are not dominated by gravitational forces and thus may adopt irregular shapes.

Tancredi and Favre (2008) review the geophysical criteria to separate a dwarf planet from a regular TNO. Assuming values for the density appropriate for icy bodies ($\rho{\sim}1$-2 g cm³), they find that the critical diameter is in the range of D${\sim}$200–900 km, or 450 km for $\rho{=}1.3$ g cm³, typical of the comparably sized Uranian and Neptunian satellites (see the JPL "Planetary Satellite Physical Parameters" page at https://ssd.jpl.nasa.gov/?sat_phys_par) and compatible with the small sample of measured TNO densities (e.g., Bannister et al., 2019). The density criterion, although it is a rough estimation, provides enough information to construct a list of candidate dwarf planets (CDPs). Tancredi and Favre (2008) obtain a list of 46 candidates containing well-known large TNOs such as Orcus, Sedna, Quaoar, Varuna, Ixion, 2003 $AZ_{84}$, 2004 $GV_9$, and 2002 $AW_{197}$. Publication of the partial results of the *Herschel* Key Program "TNOs Are Cool," the largest database of TNO size estimates, motivated a revision of this list in Pinilla-Alonso (2016). That list contained 21 CDPs of TNOs with known sizes (25 if considering the 4 already accepted by the IAU).

In this work, we adopt the Tancredi and Favre (2008) diameter criterion (step 1 in their decision tree) for classifying TNO dwarf planets. We also use the final list of sizes measured by "TNOs Are Cool," (Müller et al. 2019). Building on the preliminary work presented in Pinilla-Alonso (2016), we consider the error bars ($\sigma_D$) associated with the measured diameters. We include all the objects that have D > 450 km, even when considering the error bar (D<450 km and D+$\sigma_D$> 450 km or D > 450 km and D-$\sigma_D$< 450 km)

The list of CDPs is presented in Table 1, with 40 objects (including the 4 dwarf planets currently defined by the IAU). This list is not meant to be a definition of which objects are or are not dwarf planets–it is only a starting point to study the physical properties of TNOs large enough to be considered as such. There are differences between our list and the list of Tancredi and Favre (2008) for two reasons: First, when Tancredi and Favre (2008) was published, knowledge of the size of TNOs was sparse and most of the diameter estimates were based on absolute magnitude (e.g., Harris, 1998) rather than thermal measurements. Second, we do not follow the full decision tree presented in Tancredi and Favre (2008) as they generally require broad assumptions about density, surface roughness, and axis ratios, all of which are poorly constrained for a majority of TNOs.

Table 1: List of TNO candidate dwarf planets (CDPs)

| DP | CDP (36 TNOs) | |
|---|---|---|
| IAU accepted | D > 900 km | 450 < D < 900 km |
| Pluto, Eris, Makemake, Haumea | 2007 OR$_{10}$, Quaoar, Orcus, 2002 MS$_4$, Sedna, Salacia, | Varda, 2002 AW$_{197}$, 2003 AZ$_{84}$, 2002 UX$_{25}$, 2004 GV$_9$, 2005 RN$_{43}$, 2003 UZ$_{413}$, Varuna, Ixion, Chaos, 2007 UK$_{126}$, 2002 TC$_{302}$, 2002 XW$_{93}$, 2002 XV$_{93}$, 2003 VS$_2$, 2004 TY$_{364}$, Dziewanna, 2005 QU$_{182}$, 2005 UQ$_{513}$, 2005 TB$_{190}$, 2004 NT$_{33}$, 2004 PF$_{115}$, 2004 XA$_{192}$, 2003 FY$_{128}$, 2003 QW$_{90}$, 2002 GJ$_{32}$, 2002 KX$_{14}$, 2002 VR$_{128}$, 2001 KA$_{77}$, Huya |
| 4 | 6 | 30 |

It is worth mentioning that this list of CDPs could change with additional estimations of the size of TNO. Unfortunately, a large increase on the TNOs with well-known size is unlikely to happen in the next decade based on the current timeline of NASA and ESA missions, as there are no confirmed plans for a successor to the *Herschel Space Observatory*. *Herschel* operated in the far-infrared (55-672 μm), which was ideal for measuring peak thermal emission of TNOs between ~70-100 μm (e.g., Stansberry et al., 2008). Thus, we must instead rely on TNO stellar occultations (Ortiz et al., 2019) for the foreseeable future to estimate diameters for other TNOs. However, these studies have been limited due to the uncertain knowledge of TNO orbits and the large amount of resources required to carry out these observations.

## 3    Surface compositions of DPs and CDPs

In terms of composition, extensive studies of the primitive small bodies in the solar system (asteroids, comets, and TNOs), interstellar particles, and stellar formation regions support the idea that the primordial disk that gave birth to the planets and small bodies was formed of volatile ices, macromolecular carbonaceous species, and refractory rock (e.g., McKinnon et al., 2017). TNOs are a compositionally diverse population that includes some of the most primitive bodies in the solar system, preserving a record of the composition of the planetary disk in these distant regions (e.g., de Leon et al., 2018). It is therefore not surprising that the surface compositions of TNOs, studied by means of models of the reflectance from their surfaces, have shown that they are a mélange of amorphous silicates, complex and red carbonaceous materials, and ultra-processed amorphous carbon, with some presence of water ice and hints of other ices such as methanol (e.g., Cruikshank 2005; Barucci et al., 2011; Brown, 2012a; Barucci et al., 2019).

The size of a TNO and its surface temperature are the primary factors impacting an object's surface composition. Models of volatile retention on TNOs take these properties into consideration when used to evaluate whether or not a TNO should retain its initial volatile inventory over the age of the solar system (Schaller and Brown 2007a; Johnson et al., 2015). It is clear from these models that the dominant factor for volatile retention is diameter, which is a stand-in for surface gravity, assuming comparable densities across the TNO population. These models predict that only the four currently recognized DPs, along with a small handful of the largest CDPs, are capable of retaining a detectable amount of their original inventory of volatile ices, including $CH_4$ (methane), $N_2$, and CO. This is confirmed by the abundant number of near-infrared (NIR) spectroscopic data (0.7 to 2.2 μm) collected in the last ~15 years (e.g., Barkume et al., 2008; Guilbert et al., 2009; Barucci et al., 2011; Brown et al., 2012b).

The spectra of Pluto, Eris, and Makemake all exhibit strong $CH_4$ absorption features across the NIR spectral region (e.g., Cruikshank et al., 1976; Brown et al., 2005b; Licandro et al.; 2006[a]; Lorenzi et al., 2015). $N_2$ ice is definitively detected on Pluto (e.g., Owen et al., 1993) and inferred on Eris and Makemake through measurement of $CH_4$ band shifts (Licandro et al. 2006b; Tegler et al. 2008, 2010; Alvarez-Candal et al. 2011). CO has only been detected on Pluto to this point (e.g., Owen et al., 1993), due to strong overlapping $CH_4$ absorption features. The presence of an atmosphere around Pluto is firmly established (e.g., Elliot et al., 1989), and the presence of volatile $CH_4$ and $N_2$ on Eris and Makemake make them good candidates to support an atmosphere, at least during part of their orbits (Young and McKinnon, 2013; Hofgartner et al., 2018). The lack of detectable atmospheres in the present-day combined with these objects' high albedos provides very strong evidence for ongoing resurfacing on both of these dwarf planets (Sicardy et al., 2011; Ortiz et al., 2012).

The fourth trans-Neptunian DP, Haumea, is quite peculiar in terms of surface composition. Haumea's near-infrared spectrum shows that its surface is composed of almost pure water ice, with a mixture of both, the crystalline and amorphous phases (Brown et al., 2007, Pinilla-Alonso et al. 2009). The presence of water ice on a DP is not peculiar, having also been detected on Pluto by New Horizons (Grundy et al., 2016), but the fact that other ice species are limited to less than 8% of the surface composition is unique (Pinilla-Alonso et al., 2009). This composition is shared only with a group of smaller objects with similar orbital parameters to Haumea (Pinilla-Alonso et al. 2007, 2008), which led to the conclusion that these TNOs make up a collisional family. No other members of the Haumea collisional family are considered CDPs by our criteria. The lack of detected volatile ices on Haumea may be due to the collision that formed the family, and the subsequent thermal heating that Haumea underwent.

Volatile retention models indicate that a limited number of CDPs, specifically 2007 $OR_{10}$, Quaoar, and Sedna may retain some volatile ices on their surfaces (Schaller and Brown, 2007a; Brown, 2012a). Indeed, $CH_4$, the least volatile of the volatile ices (e.g., Fray and Schmitt, 2009), has been suggested on all three of these objects, with $N_2$ also posited on Sedna (Barucci et al., 2005; Schaller and Brown, 2007b; Emery et al. 2007). As shown in Figure 1, the visible spectral slopes of these CDPs indicate their surfaces are significantly redder than those of the DPs, possibly due to the widespread presence of tholins. This implies a different equilibrium between the mechanisms altering the surface ices (Gil-Hutton, 2002) similar to what is inferred on the volatile-dominated DPs.

In the next size tier of CDPs (all with D > 900 km) are Orcus, 2002 $MS_4$, and Salacia. The visible colors of these objects are neutral and their geometric albedos are relatively low (Fig. 1), possibly indicating surfaces dominated by amorphous carbon and lacking volatile ices. However, the water-dominated surface of Orcus also may show evidence of $CH_4$, $NH_3$, and their irradiation products (Delsanti et al., 2010; Carry et al., 2011), though this is not confirmed.

If we extend the study of albedo to the rest of the CDPs, this group coincides with what is referred to in Bannister et al. (2019) as "mid-sized TNOs." The distribution of albedo vs. diameter for this group is more similar to that of the TNO population as a whole (Fig. 1, left panel), which is consistent with volatile-free surfaces. For these smaller objects, according to VIS/NIR spectroscopy, water ice is the most abundant ice, though it never dominates the surface. On the contrary, it appears mixed with other materials, typically featureless in the VIS/NIR, such as silicates and complex organics (tholins). It is therefore surprising that the smaller TNOs (D≲450km) have a range of albedos from 2% to 30%, independent of their size. In the case of the CDPs smaller than 900 km in size, the albedo tends to decrease with an increase in size.

The comparative study of the visible colors of TNOs and CDPs shows that the majority of CDPs have red surfaces (S'>10%/1000 Å; Fig. 1, right panel). There is a notable group of neutral CDPs, including Orcus, but this group is smaller than the red group. An important feature of the right panel of Figure 1 is the near lack of CDPs with spectral slopes between 6.6%/1000 Å (2003 UZ$_{413}$, Peixinho et al. 2015) and 17.1%/1000 Å (Varda, Peixinho et al. 2015); only Salacia, with a spectral slope of 12.6%/1000 Å, occupies this region (Pinilla-Alonso et al., 2008).

[PLACE FIGURE 1 (double panel) HERE]

As shown in Figure 1, the albedo and colors of the CDPs have some peculiarities that may be associated with the properties of their surface material. Lying in diameter between the geologically active dwarf planets and the collisionally evolved small bodies, the CDPs represent the best tracers for the composition of the solar nebula. Improving compositional information about these objects will help to establish clearer connections between their surface compositions and their physical parameters (e.g., size, albedo, dynamical class). These connections will, in turn, better constrain models for the formation and evolution of the solar system, and act as a model for planetary systems discovered around other stars. The biggest potential to extend this knowledge in the near future comes from gathering data at wavelengths longer than 2.2 μm, where ices such as $CH_4$, $N_2$, CO, $H_2O$, and $CO_2$ have their fundamental absorptions. This is also a wavelength regime where non-methane hydrocarbons (ethane, ethylene, acetylene, propane, etc.) show very distinctive absorptions. Outstanding questions pertaining to DPs and CDPs include:

- Can we obtain definitive proof of the presence of the volatile ices $N_2$ and CO on Makemake and Eris, and if so, what does this mean for the possibility of atmospheres around these dwarf planets?
- What is the inventory and nature of non-methane hydrocarbons and complex organics (tholins) on DPs and CDPs, and what can comparison of the two populations reveal about the chemical evolution of TNO surfaces?
- What, if any, volatiles are present on the largest CDPs?
- What ice species, if any, are present on the smallest TNOs, and what does this reveal about their formation environment?

The remainder of this paper is dedicated to a description of NASA's next great observatory, the *James Webb Space Telescope* (*JWST*), which will help address the above questions with its large aperture, sensitive instrumentation, and expanded wavelength coverage.

## 4    Potential of the James Webb Space Telescope

The *James Webb Space Telescope* offers significant promise for characterizing the compositions of dwarf planets and other TNOs in detail, and for a significant sample, for the first time. Its wavelength coverage (0.6–28 μm), exceptional sensitivity, unprecedented spatial resolution, and comprehensive suite of modest resolution spectroscopic instrumentation could enable breakthroughs in linking composition to dynamical classification. This is particularly true in the critical 1–5 μm region, where many molecules of interest have numerous absorption bands. *JWST* will also have the spectral sensitivity and resolving power in the 5–10 μm range to characterize organic molecules at wavelengths that have previously been impossible, and will offer improvements in imaging sensitivity near 25 μm that could prove powerful for characterizing TNO albedos, temperatures, and sizes, particularly when combined with longer wavelength observations from Herschel (70–500 μm) and ALMA (>800 μm).

*JWST* is primarily designed as an astrophysical observatory covering wavelengths from 0.6–28 μm, but will also provide ground-breaking capabilities for solar system science, particularly for studies of TNOs. These capabilities result from the large (6.5-meter effective diameter) primary mirror which, like the near-infrared instruments onboard, is passively cooled to around 40 K, as well as the significantly more complex and modern instrumentation when compared, for example, to those on *Spitzer*. There is no overlap in capabilities with *Herschel*. The observatory and science goals are described in Gardner et al. (2006). Solar system science capabilities are described in considerable detail in Milam et al. (2016). Ten companion papers are in the same volume of *Publications of the Astronomical Society of the Pacific*, including Parker et al. (2016), which focuses on TNO observations. Here we provide a brief summary, focused on TNO science applications. *JWST* is currently in the late stages of integration, with the instruments and telescope already assembled into a single subsystem and the spacecraft subsystem in the final stages of test. Launch is now expected in March 2021. Science operations will be supported by the Space Telescope Science Institute (STScI), which also operates the *Hubble Space Telescope*.

## 4.1 Observatory and ground system capabilities

### 4.1.1 Orbit, field of regard, and moving-target tracking

*JWST*, like the *Spitzer* and *Herschel* observatories, will operate at the Earth-Sun L2 point, approximately 0.01 AU, external to Earth's solar orbit. The observatory telescope and instruments are passively cooled by means of a large sunshade; in order to keep those components shaded, the observatory is restricted to point between 85° and 135° in solar elongation angle (Sun-*JWST*-Target angle). Note that observations at, or even near, opposition cannot be made. That limitation, combined with the fact that the observatory can be pointed to any azimuth around the Sun-observatory vector, defines the instantaneous "field of regard" (FOR). The FOR is thus an annulus of one celestial sphere, with two 50°-wide regions centered on the ecliptic plane. The range of roll angles about the boresight is only ±5°; for observations near the ecliptic, the available on-sky orientation is thus also limited to the same range (at higher ecliptic latitudes wider ranges of orientation can be accessed, depending on the epoch of the observation).

The observatory is required to be able to track moving targets at rates up to 108″/hr (30 milliarcseconds/second, the maximum apparent rate of Mars as seen from L2), so it is more than adequate for observations of all TNOs and Centaurs (Milam et al., 2016). The length of science exposures is limited by the time a guide star remains in the field of view of the guider. For moving-target observations the effective FOV of the guider is 2.0′ × 2.0′. At the maximum track rate, the resulting limit on exposure time would thus be about 2000 seconds. For observations of Centaurs and TNOs this guide-star limit on exposure time is superseded by the requirement that individual exposures be ≤10,000 seconds. Beyond that, observers must specify multiple exposures if additional time is needed.

### 4.1.2   Observation planning and documentation

Extensive on-line documentation for the *JWST* observatory, instrumentation, and planning tools is easily accessible at the STScI website: https://jwst-docs.stsci.edu/. High-level topics covered there are: calls for proposals (including science policies and a description of the guaranteed-time observation programs), proposal planning (including observatory performance and constraints), proposing tools (Exposure Time Calculator, Astronomer's Proposal Tool, visibility tools), instrumentation, and data products. Links to the planning tools themselves are also available at the same website. The documentation currently includes about 15 separate pages giving specific guidance for planning observations of moving targets, including details of how to use the Astronomer's Proposal Tool (APT), Exposure Time Calculator (ETC), and Moving-target Visibility Tool (MTVT).

*JWST* observations are defined using observing templates in APT, the same tool used to define observations using *Hubble*. The *JWST* templates are similar in function to the templates used for the *Spitzer* and *Herschel* observatories. They provide users with a fairly intuitive workflow for defining an observation, while helping to avoid making choices for instrument parameters that might negatively impact data quality. All templates for all instruments support observations of moving targets, with only a few minor restrictions relative to capabilities for fixed targets. An example of such a restriction is that target acquisition must be performed on the moving target itself, while for fixed targets acquisition on an offset target is supported.

Signal-to-noise calculations are performed using the *JWST* ETC (jwst.etc.stsci.edu). At present the ETC is not well-suited for defining a solar system object and computing the SNR that would result from an observation of it on a date or over a range of dates. However, users can create a model spectrum using an appropriately normalized "solar" (G2V) stellar spectrum, and, if needed, combine that with a similarly normalized blackbody spectrum. For TNOs, whose observing circumstances change very little over a year-long observing cycle, this approach is adequate. Users can also compute a model spectrum separately and upload that into the ETC. The left panel of Figure 2 shows an example ETC calculation using a 1000-second exposure with the NIRSpec integral field unit (IFU) and the low-resolution prism to observe TNO (55565) 2002 AW$_{197}$, where the spectrum was modeled as described above.

Visibility of targets from *JWST* can be calculated using the JPL Horizons system or the Python package jwst_mtvt (documentation and code available via the URL above). In Horizons, specify "@jwst" as the Observer Location and in the Table Settings limit the solar elongation angle to 85°–135°. The jwst_mtvt package generates tabular and graphical output giving the dates when a desired target is within that elongation range, and provides the range of on-sky orientation angles of the *JWST* focal plane within those observability windows. (The orientation angle is generally not of great interest for observations of isolated sources such as TNOs, but can be critical for planetary satellites.) The right panel of Figure 2 shows the graphical observability summary for (55565) 2002 AW$_{197}$ between January 2021 and July 2022.

PLACE FIGURE 2 HERE (two panels)

### 4.1.3 Pointing accuracy and target ephemerides

The pointing accuracy of *JWST* is expected to be between 0.3-0.45″ (1-σ), depending on the distance between the science aperture and the guide star in the Fine Guidance Sensor (FGS). The pointing stability for moving targets over a 1000-second exposure is estimated to be between 6.2-6.7 mas (1-σ). The pointing accuracy rules out blind pointing for placement of targets in the NIRSpec fixed slits (§4.2.2) and the MIRI Low-Resolution Spectrometer (§4.2.3), regardless of the quality of the target ephemeris. Target acquisition (TA) will be required to accurately place targets in slits. For targets with more uncertain ephemerides (~1″), TA may be required to place targets in the MIRI Medium-Resolution Spectrometer and NIRSpec IFU apertures. Targets with ephemeris uncertainties of a few arcsecs or more, are less likely to be targetable with *JWST*, given that they must first be blindly placed in the TA aperture or region of interest. Reporting additional astrometry of these targets to the Minor Planet Center (MPC) is recommended prior to proposing for spectroscopic observations with *JWST*.

## 4.2 Instrumentation

*JWST* has four science instruments providing imaging (NIRCam, NIRISS, and MIRI) and spectroscopy (the previous three and NIRSpec) covering wavelengths from 0.6–28 microns. The performance of the instruments is described in more detail below. The key modes for TNO science are likely to be NIRSpec IFU spectroscopy (0.7–5 μm), NIRCam imaging (0.7–5 μm), and MIRI imaging and spectroscopy (5–28 μm). The NIRSpec slitted spectroscopy mode could become important, but due to the small slit widths may require additional time to fully commission for moving targets.

All of the instruments utilize the "sample-up-the-ramp" approach for reading out the detectors. The 3 NIR instruments use 2048 × 2048 HgCdTe detector arrays while MIRI uses 1024 × 1024 Si:As arrays. By sampling the signal repeatedly and non-destructively as charge collects during an exposure, cosmic ray strikes can be identified during the fitting process that converts the data ramps to slope images. Additional benefits include better calibration of non-linear effects and saturation of the detectors, as well as improved sensitivity. While cosmic rays will be detected and corrected at the individual integration ramp level, dithering will still be necessary to allow rejection of the fainter cosmic rays and for bad-pixel replacement in the final data products.

### 4.2.1 NIRCam

The Near-IR Camera (NIRCam) is the primary imager for *JWST*, with a total field of view (FOV) of 9.5 square-arcmins and providing simultaneous 2-filter imaging in separate 0.7–2.3 μm and 2.4–4.8 μm channels. The instrument consists of two fully redundant modules each with one short-wavelength (SW) and one long-wavelength (LW) channel. The four focal planes are composed of 10 2048 × 2048 HgCdTe detector arrays, four in the shorter wavelength channels and two in the longer wavelength channels of both modules. For observations of specific TNOs observers may prefer to use a single module (with a 2.2′ × 2.2′ FOV). For surveys both modules can be used simultaneously.

NIRCam has 13 SW and 16 LW filters. Their spectral widths fall into "wide," "medium," and "narrow" categories, with approximate fractional bandpasses of 25%, 10%, and 1%, respectively. Two ultra-wide filters at 1.5 μm and 3.22 μm will be particularly useful for detecting the faintest TNOs, e.g., in surveys.

The *JWST* telescope will provide diffraction-limited performance at wavelengths >2 µm, but image quality in terms of the full-width at half maximum (FWHM) of the point spread function (PSF) is nearly diffraction-limited through even the shortest filter at 0.7 µm. However, the NIRCam pixel-scale in the SW and LW channels are 31 mas and 63 mas, set to provide Nyquist spatial sampling of the PSF at 2 µm and 4 µm, respectively. In order to take advantage of the full spatial resolution at shorter wavelengths, observers will need to include dithers in their observations. It is worth noting that, at the shortest wavelengths, the spatial resolution available with NIRCam will be approximately two times better than the best available with the *Hubble Space Telescope* and its WFC3 instrument.

With *JWST's* large aperture and modern detectors, NIRCam will provide significantly higher sensitivity than *Hubble* from 0.9–1.8 µm. Due to the IR-optimized throughput of *JWST*, however, the sensitivity at 0.7 µm is nearly the same as for *Hubble*. Between 1.8 µm and 5 µm Hubble has no capability, and observatories such as *Spitzer* and *WISE* are orders of magnitude less sensitive. Figure 3 compares the 10-σ sensitivity of NIRCam imaging with 1000-second exposures to hypothetical spectral energy distributions of TNOs with a range of compositions (see §4.2.2.1 for more details of the spectral models). Where the spectra lie above the NIRCam sensitivity values signal to noise (SNR) will exceed 10; where the spectra fall below the sensitivity value SNR will be correspondingly lower. The figure illustrates that near-IR color photometry of small (~50-km diameter) TNOs in the cold-classical population will be possible in a modest amount of observing time, as will L-band (and shorter) characterization of Centaurs at the distance of Neptune. For brighter objects, observers may consider acquiring spectra with NIRSpec (see below) rather than obtaining NIRCam colors.

NIRCam has also great potential for performing surveys for faint TNOs. In particular, through the F150W2 filter it should be possible to detect objects at $m_v=27$ (equivalent to a 35-km diameter object at 45 AU, comparable to (486958) 2014 MU$_{69}$) at an SNR of five in a 100-second exposure (short enough that typical TNOs will not be appreciably trailed in a fixed-pointing image). Using digital tracking techniques, it should be possible to push significantly fainter than that.

[ PLACE FIGURE 3 HERE ]

### 4.2.2  NIRSpec

The Near-IR Spectrometer (NIRSpec) is the 0.6–5 µm spectrometer for *JWST* and offers imaging spectroscopy (via an integral field unit, or IFU), and fixed-slit and multi-object spectroscopy. Due to the small widths of the fixed slits (200 mas), the IFU with its $3'' \times 3''$ FOV will be more forgiving for observations of targets with any appreciable ephemeris uncertainty (see §4.1.3). This is especially true because target acquisition for moving targets must be done using a small $1.6'' \times 1.6''$ aperture. This limitation of NIRSpec will require that target ephemerides be exceptionally well-known prior to scheduling.

NIRSpec utilizes a pair of the same H2RG detectors as NIRCam. Dispersers can be used in combination with the IFU or slits, and span wavelengths from 0.6–5.3 µm. The dispersers consist of a prism requiring a single observation to cover all wavelengths (resolving power 30R < 300), three medium-resolution (R≈1000), and three high-resolution (R≈2700) gratings, with the gratings requiring three separate observations to span the full wavelength range. When the high-resolution gratings are used with the IFU or fixed slits, small gaps in spectral coverage result from the physical gap between the two detectors; for the PRISM and medium-resolution gratings the spectra fall entirely on a single detector.

Figure 3 compares the 10-σ sensitivity of NIRSpec IFU spectroscopy using 1000-second exposures to hypothetical spectral energy distributions of TNOs with a range of compositions. Sensitivity for the PRISM and medium- and high-resolution gratings is shown. While spectroscopy using the slits will be more sensitive, here we focus just on the IFU due to possible difficulties with placing moving targets into those slits, as mentioned earlier. The spectral models are based on albedo spectra for the objects named in the figure legend. For each object Hapke models (Hapke, 2012) were used to fit available data in the visible – near-IR range, and extended to 5 μm. The fitting included matching the known geometric albedo of each object. Those albedo spectra were then used to predict spectral energy distributions for hypothetical objects with sizes and distances not necessarily representative of the original objects. These hypothetical objects allow us to show the range of spectral features seen for TNOs, the dynamic range within the spectrum for single objects with different compositions, and the range of overall brightness to be expected for TNOs with a range of sizes and distances. Pluto is somewhat of a special case, as it is shown for its actual size and the distance appropriate for the end of the *JWST* mission, c. 2032. Additional details on NIRSpec observations of TNOs can be found in Métayer et al. (2019)

The NIRSpec IFU pixel scale is 100 mas, so the PSF (comparable to that of NIRCam) is under-sampled at all wavelengths. Fixed slits are ≈3.5″ long and either 0.2″ or 0.4″ wide. A larger 1.6″ × 1.6″ aperture is used for acquiring single targets and can also be used for spectroscopy. As mentioned earlier, the small sizes of these slits place tight requirements on ephemeris accuracy, and observers will be required to show that ephemerides for their targets are sufficient prior to observations being scheduled. Spectra from the slits fall on a region of the detectors that is behind an opaque mask, minimizing background light and offering higher sensitivity than is possible with the IFU.

The NIRSpec IFU offers imaging spectroscopy with a 3″ × 3″ field of view. Spaxels are 0.1″ square, matched to the pixel scale. Dithers can be used to improve spatial sampling of the PSF if desired. Due to its larger aperture, observations with the IFU are less sensitive to target ephemeris uncertainties, and the location of the target within the scene can be determined after the fact to enable optimal spectral extraction. The IFU also simultaneously characterizes the background around the target in two dimensions, which is an advantage compared to the slits. However, spectra from the IFU fall on portions of the detectors that are masked by the micro-shutter array (MSA). The MSA is not perfectly opaque even for closed shutters, and there are a number of failed-open shutters that will allow dispersed light from objects and the zodi to fall on the detectors during IFU exposures. For these reasons the IFU is somewhat less sensitive than the slits, and dithers are strongly encouraged in order to help remove the effects of any sources that may fall on open shutters.

### 4.2.3 MIRI

The Mid-IR Instrument (MIRI) offers both imaging and spectroscopy in the 5–28 μm wavelength range. The imager uses a single 1024 × 1024 pixel detector with a pixel scale of 0.11″ and a FOV of 74″ × 113″ (a portion of the detector is dedicated to coronagraphic imaging). The PSF is diffraction-limited at all wavelengths and is slightly under-sampled by the pixels at the shorter wavelengths (FWHM = 0.18″ at 5.6 μm) and highly oversampled at the longest (FWHM = 0.82″ at 25.5 μm). Figure 4 shows the total system response through the nine MIRI filters. Images are acquired through a single filter at a time (in contrast to NIRCam, in which data is collected through two filters simultaneously).

MIRI also offers medium resolving-power imaging spectroscopy from 4.9–28.8 μm via the medium resolution spectrometer (MRS), which is an IFU similar to the one in NIRSpec. Light is dispersed via gratings, resulting in spectral resolving power ranging from approximately 1400 at the longest wavelengths to about 3500 at the shortest wavelengths. In a single exposure, spectra are acquired in each of the four wavelength channels, two spectral channels being dispersed onto separate parts of each of the two MRS detectors. The four channels span wavelengths of 4.9–7.7 μm, 7.5–11.7 μm, 11.5–18.1 μm, and 17.7–28.8 μm, respectively. For a single grating setting the spectrum spans a wavelength band that is about one third of the full wavelength range covered by each of the 4 channels. Continuous spectral coverage requires taking three exposures, each using a different grating setting. Figure 4 shows the 12 MRS spectral sections (four channels, and three bands within each channel). The four MRS wavelength channels have fields of view that overlap on-sky, with footprints ranging from $3.3'' \times 3.7''$ in the shortest wavelength channel to $7.2'' \times 7.9''$ in the longest wavelength channel.

Finally, the MIRI low-resolution spectrometer (LRS) provides spectra spanning 5 μm to slightly beyond 10 μm in a single exposure (see Figure 4). The light is dispersed by a prism and the resolving power ranges from 40–160 over that wavelength range. The spectrum is dispersed onto the same detector used for MIRI imaging (separate from the two MRS detectors), and can be taken through a $0.5'' \times 4.7''$ slit, or slitless (the latter intended primarily for observations of exoplanet transits).

[PLACE FIGURE 4 HERE ]

Figure 4 illustrates the sensitivity of MIRI in both imaging and spectroscopic modes. Generally, MIRI will provide much higher sensitivity than the *Spitzer* infrared spectrograph (IRS) instrument, hardly surprising given the huge disparity between the *JWST* and *Spitzer* apertures. However, beyond about 20 μm MIRI sensitivity is limited by thermal emission from the much warmer telescope (*JWST* operates at around 40 K while *Spitzer* was typically at around 15 K), and the sensitivity gain of *JWST*/MIRI over *Spitzer* is more modest. The much smaller PSF of *JWST* does, however, significantly reduce effective noise caused by background sources (referred to as confusion noise) in the scene relative to *Spitzer*. Confusion is primarily caused by thermal emission from dust in distant, un-resolved galaxies, and is more important at MIRI wavelengths than in the VIS and NIR. In the context of Figure 4, Pluto provides a concrete example for which the *JWST*/MIRI sensitivity can be compared to that of *Spitzer*. MIRI imaging at 25.5 μm should detect Pluto with an SNR of ~50 in a 500 second exposure; *Spitzer* 500 sec observations at 24 μm gave an SNR of 25 (Lellouch et al., 2011). For MIRI R≈1500 spectroscopy of Pluto only (Figure 4 does not include the contribution of Charon) using 1000 sec exposures should give SNR≈2 per spectral element, over wavelengths of 15–28 μm if the data are un-binned. If the data are binned by 15 spectral elements (giving R≈100), the resulting spectrum would have SNR≈8. *Spitzer* 720 sec spectra of Pluto (including Charon) with R≈90 also gave SNR≈8 (Lellouch et al. 2011), but only over wavelengths of about 22–36 μm.

The 6.5 m primary mirror of *JWST* will enable much higher spatial resolution than any previous space observatories with instruments covering similar wavelengths, which all had ≤1 m primary mirrors (e.g. *IRAS*, *ISO*, *WISE*, *Spitzer*, *AKARI*). The FWHM of the MIRI PSF ranges from $0.22''$–$0.82''$ from 5.6–28.5 μm. In imaging mode, the pixel scale ($0.11''$) is such that the PSF is Nyquist sampled at the short wavelengths, and highly over-sampled at the longer wavelengths. For the medium-resolution imaging spectrometer (MRS) the pixel scales in the 4 spectral channels mentioned above are 0.176″, 0.277″, 0.387″ and 0.645″, respectively, with the result that the spectral PSF is somewhat under-sampled by the pixels at all MRS wavelengths. This significant leap in spatial resolution for a spaced-based mid-IR instrument will be particularly important for Jupiter, Saturn, and their large satellites. In the Kuiper belt, however, the primary benefit may be in better detection and removal of flux from background sources. A few of the most well-separated binary systems (e.g., Pluto/Charon) can be resolved by MIRI at the shorter wavelengths where reflected light tends to dominate TNO spectra (see Figure 4); at the longer wavelengths, where the thermal emission is brightest, even Pluto and Charon will be blended.

### 4.2.4 NIRISS

The Near-IR Imager and Slitless Spectrograph (NIRISS) offers a unique interferometric imaging mode (aperture masking interferometry, or AMI) in four filters spanning 2.5–5 μm. AMI delivers spatial resolution roughly 2.4× better (i.e. the PSF is that much sharper) than NIRCam imaging at those wavelengths. This capability relies on a pupil mask with seven small sub-apertures within the extent of the *JWST* primary mirror, where each pair-wise baseline vector is unique. Even though the pupil mask greatly reduces the throughput, the NIRISS optics are otherwise very efficient and in AMI mode NIRISS is only a factor of five less sensitive than NIRCam imaging. For $m_v=23$ TNO with a neutral spectrum, NIRISS/AMI can achieve an SNR of around 200 in a 100 sec exposure at 2.8 μm, and would be able to resolve satellites at separations ≥0.09″. AMI could be used to measure colors of TNO binaries in the L- and M-band region for the first time.

## 4.3 Data products and archive

Plans are in place to provide better support, relative to what has been provided for *Hubble*, for *JWST* moving-target observations, both in terms of the data processing and the ability to find data in the archive. Users are now strongly encouraged to use standard designations for their moving targets when submitting their observations. The Astronomer's Proposal Tool (APT) makes it easy to do so because it now includes the capability to resolve target names and retrieve orbital elements from JPL Horizons. A secondary naming field can be used to specify details in addition to target names, if desired (e.g. "west elongation" or "longitude 90"). This means that archive searches will reliably return data for users who search on normal target names.

The *JWST* data pipeline is also implementing the capability to co-add multiple exposures of moving targets in the frame of the target. This is critical, e.g. for improving SNR and PSF sampling by combining dithered exposures into a final image or spectrum, or producing maps of extended objects such as comets. Such data products have been available for recent missions such as *Spitzer* and *Herschel*, but have not been produced for *Hubble* observations.

## 4.4 Guaranteed-time observations (GTOs)

*JWST* includes a pre-determined allocation of time for scientists who have contributed directly to the development of the observatory, e.g., the principal investigators for each of the science instruments. Four guaranteed-time observer programs totaling about 70 hours will be focused on the TNOs and Centaurs, primarily on the dwarf planets including Pluto, Eris, Makemake, and Haumea. NIR spectroscopy with NIRSpec makes up the bulk of the time, with additional MIRI spectroscopy and MIRI imaging observations. Details regarding these programs can be found at http://www.stsci.edu/jwst/observing-programs/approved-

, including the observation specifications themselves. These programs will be finalized in March 2020. General observers wishing to make similar observations of the same targets will be required to provide a justification for the duplication, e.g., the need to study time-variable phenomena, or obtain significantly higher SNR.

## 5    Summary

Surface compositions of TNOs appear to be correlated with size, with the largest TNOs, the dwarf planets, exhibiting dynamic, volatile-dominated surfaces. We refer to the next-lowest size tier as candidate dwarf planets. These objects appear to be vastly different from the dwarf planets in terms of color, albedo, and surface composition, even though they are closer in size to the dwarf planets than the small TNOs. Fundamental questions remain about the connections between the dwarf planets, candidate dwarf planets, and small TNOs, many of which cannot be answered with current ground-based facilities and instrumentation. The *James Webb Space Telescope* will provide the higher sensitivity and extended wavelength range needed to address these outstanding issues. *JWST* has a 6.5-meter diameter primary mirror and is equipped with four science instruments with imaging and spectroscopic modes that cover 0.6-28 μm. All TNOs can be tracked with *JWST* and plans are already in place to process moving target observations. Set to launch no later than March 2021, *JWST* is poised to revolutionize the study of TNO surface compositions in the coming decade.

**Acknowledgements**

The authors thank the conveners of the scientific workshop, "The trans-Neptunian Solar System", held in the University in Coimbra, Portugal, in March 2018. The authors would also like to recognize the work of Michael S.P. Kelley on the jwst_mtvt and Michael Mommert on the astroquery adaptation for JPL Horizons.

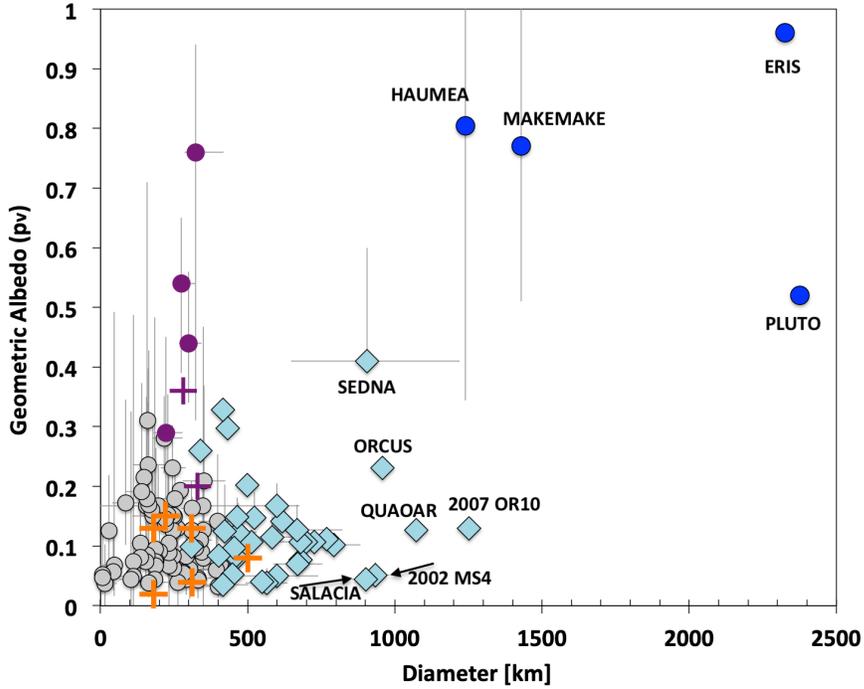

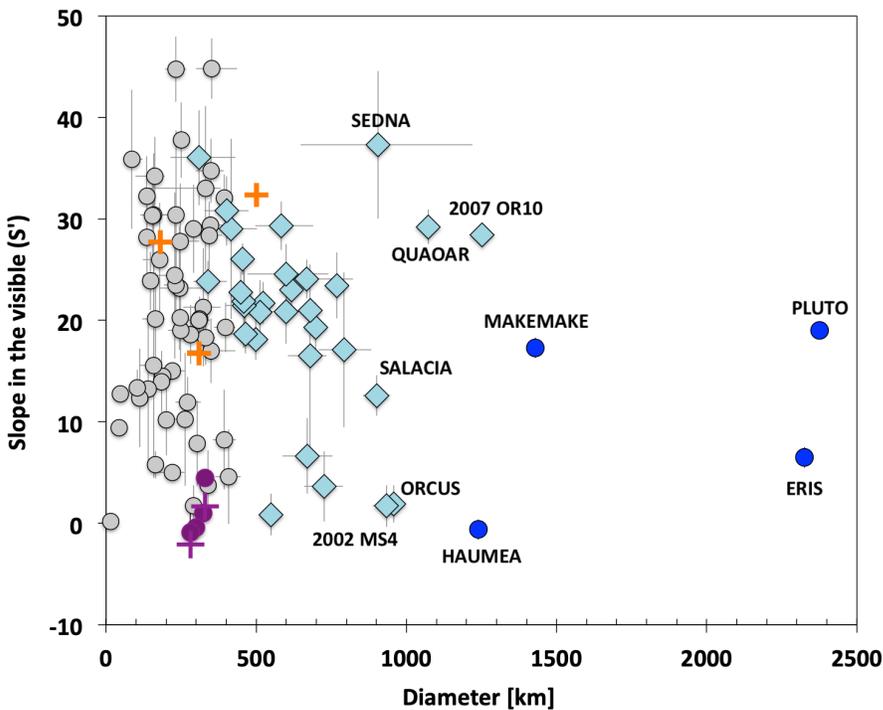

**Figure 1**. *Left*: Geometric albedo vs. diameter for a wide range of TNOs. Blue dots represent the Dps as defined by the IAU at this moment, light blue diamonds represent the CDPs, and purple dots represent the members of the Haumea family. Crosses represent TNOs with an estimation for the upper limit on diameter (lower limit in geometric albedo). Purple crosses represent members of the Haumea family with an estimation of the upper limit on diameter. Grey dots represent the TNOs with a good determination of the diameter and albedo that are not CDPs. Some well-known TNOs (D > 900 m) are labeled. The geometric albedo of the CDPs is more similar to that of medium and small TNOs than to that of the DPs. Only Sedna and some members of the Haumea family have a geometric albedo above 40%, which suggests a large amount of ice on their surfaces. *Right*: Visible spectral slope (%/1000 Å) vs. diameter for the same collection of TNOs. Note that, on average, the CDPs are redder than the DPs, potentially due to ongoing resurfacing on the DPs. Note also that lack of CDPs with inermediate slopes (6.6 < S' < 17.1) (Spectral slope data from Pinilla-Alonso et al., 2007, 2008, 2009; Lorenzi et al., 2016; Hainaut et al., 2012; Bauer et al., 2013; Peixinho et al., 2015; Szabó et al., 2018.) Diameters and geometric albedos from "TNOs are Cool," http://public-tnosarecool.lesia.obspm.fr/Published-results.html.)

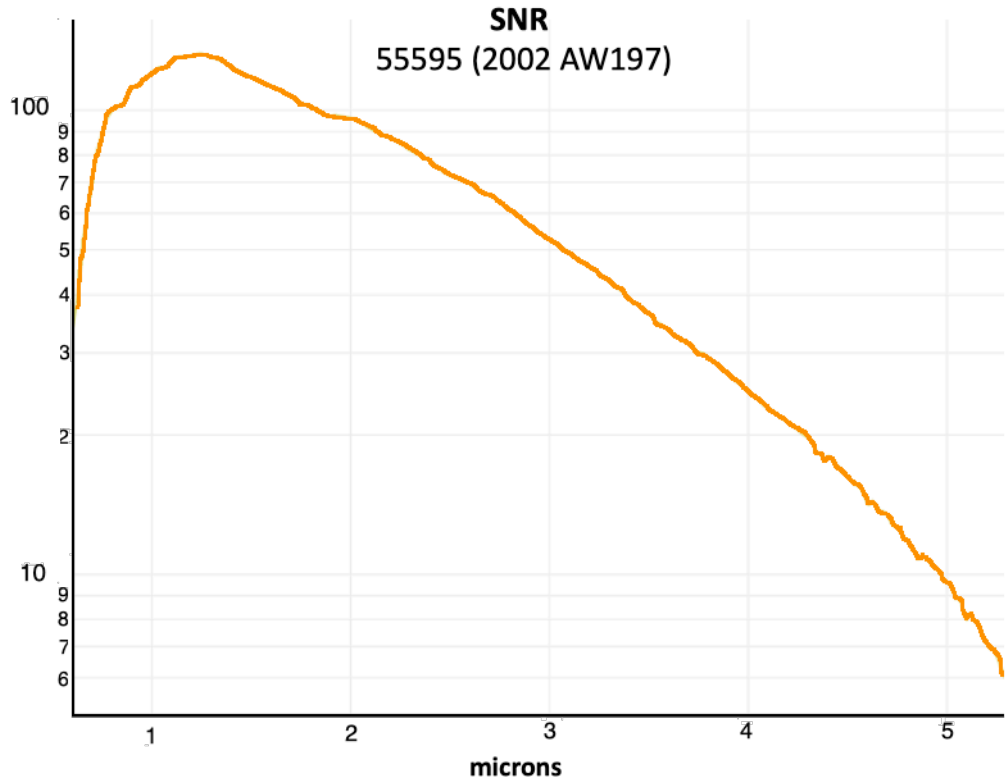

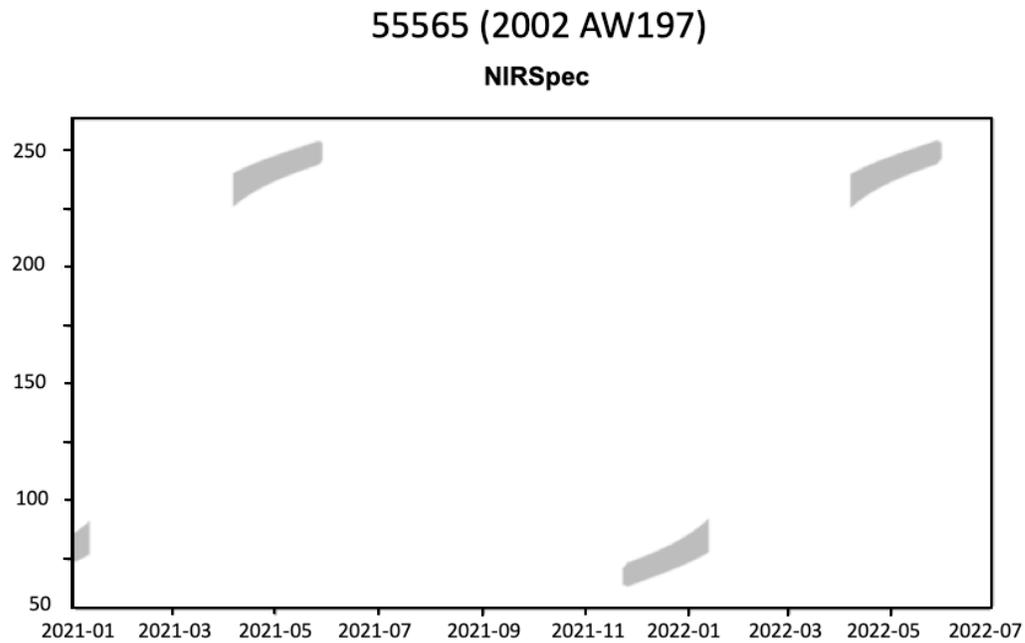

**Figure 2.** *Left*: Example *JWST* ETC calculation of signal-to-noise for a 1000-second NIRSpec exposure on target (55565) 2002 AW$_{197}$ using the integral field unit (IFU) and the PRISM. *Right*: Example output from the visibility tool, jwst_mtvt, for the same target.

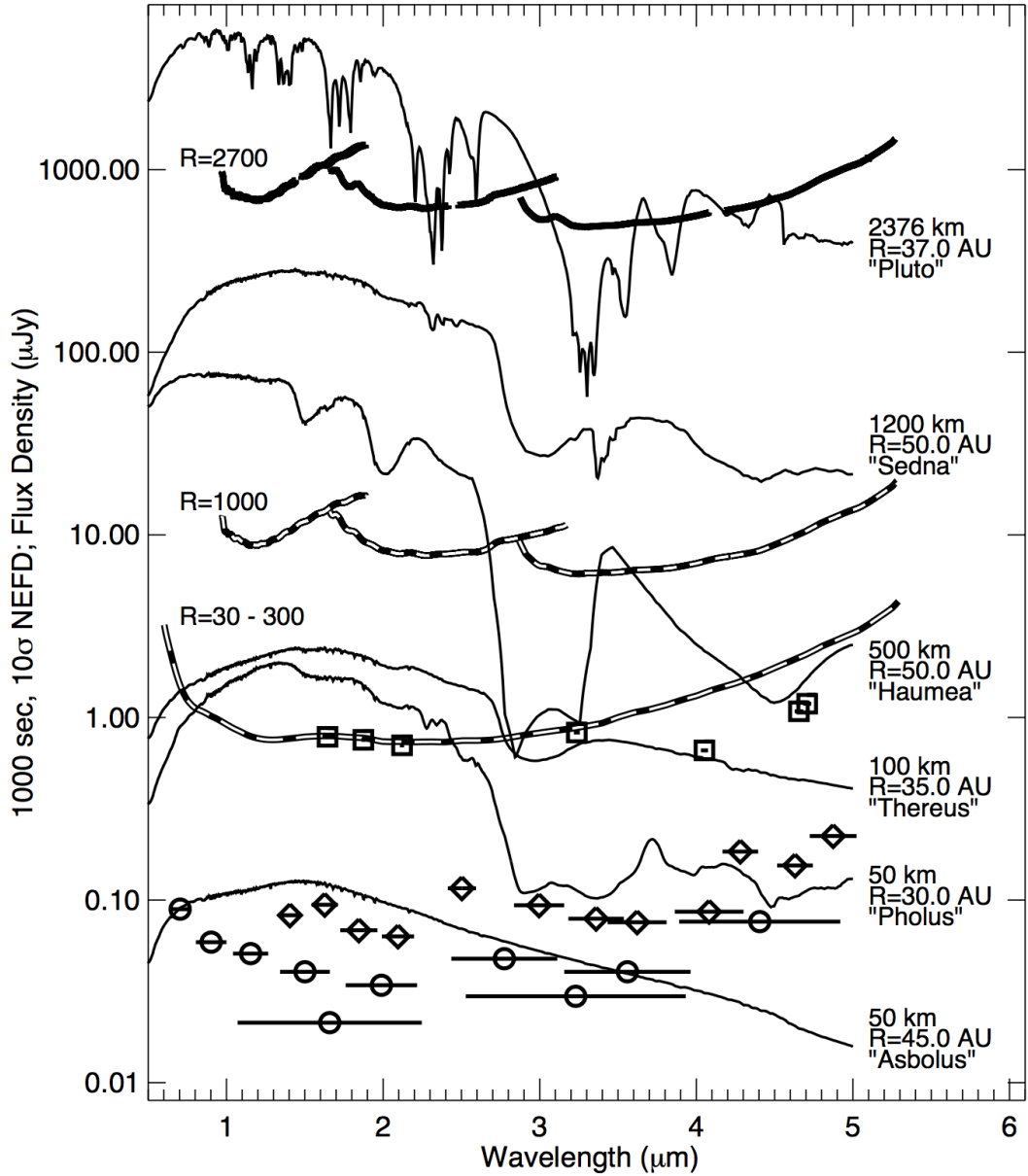

**Figure 3.** NIRCam and NIRSpec 10-σ Noise Equivalent Flux Density (NEFD, aka sensitivity) compared to spectral energy distributions of hypothetical TNOs with albedo spectra based on those of (bottom to top) Asbolus, Pholus, Thereus, Haumea, Sedna and Pluto. Albedo spectra are models fit to available VIS to NIR data for those targets. Diameters and distances of the hypothetical objects are given in the labels. NIRCam sensitivities are represented by horizontal bars indicating the filter bandpasses, with the wide, medium, and narrow filters shown as dots, diamonds and squares, respectively. NIRSpec sensitivity for the IFU is shown by dashed lines labeled as R = 30-300 (PRISM, 30<R<300), dashed line labeled as R=1000 and black thick line, R=2700 gratings. Note that the resolving power of the gratings is not constant but varies much less than for the PRISM. Spectra for the high-resolution gratings are dispersed across two detectors, resulting in small gaps in spectral coverage (clearly visible for R=2700).

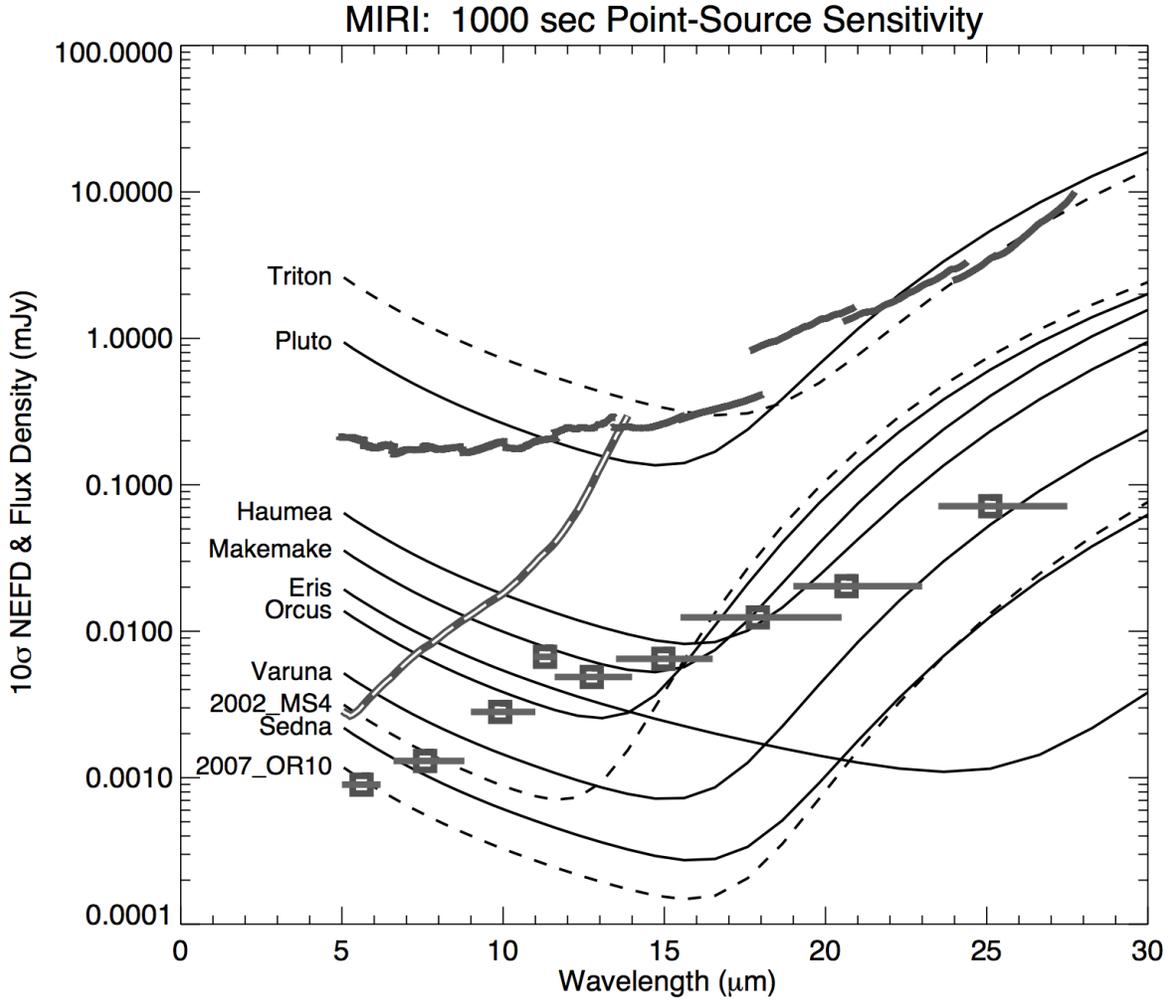

**Figure 4.** MIRI 10-σ Noise Equivalent Flux Density (NEFD, aka sensitivity) compared to spectral energy distributions of various TNOs. Measured geometric albedos and diameters are used to predict the reflected component (assuming constant albedo), which dominates shortward of ~15μm, and thermal emission using an assumed phase integral of 0.39. Horizontal lines with square symbols show the MIRI sensitivity in the imaging bandpasses, while the thick short curves at the top give the MRS sensitivity. The dashed-solitd curve from 5–13 μm gives the LRS sensitivity.